\documentstyle[12pt]{article}
\begin{document}

\title{Chiral symmetry breaking and violation of the Wiedemann-Franz law in underdoped cuprates}
\author{Guo-Zhu Liu$^{1,3}$ and Geng Cheng$^{2,3}$ \\
$^{1}${\small {\it Lab of Quantum Communication and Quantum Computation, }}\\
{\small {\it University of Science and Technology of China, }}\\
{\small {\it Hefei, Anhui, 230026, P.R. China }}\\
$^{2}${\small {\it CCAST (World Laboratory), P.O. Box 8730, Beijing 100080,
P.R. China }}\\
{\small {\it $^{3}$Department of Astronomy and Applied Physics, }}\\
{\small {\it University of Science and Technology of China, }}\\
{\small {\it Hefei, Anhui, 230026, P.R. China}}}
\maketitle

\begin{abstract}
\baselineskip20pt We propose that the recently observed violation of the Wiedemann-Franz law in the normal state of underdoped cuprates is caused by spin-charge separation and dynamical chiral symmetry breaking in a (2+1)-dimensional system consisting of massless Dirac fermions, charged bosons and a gauge field. While the $d$-wave spinon gap vanishes at the Fermi points, the nodal fermions acquire a finite mass due to strong gauge fluctuations. This mass provides a gap below which no free fermions can be excited. This implies that there is not a residual linear term for the thermal conductivity, in good agreement with experiments. Other physical implications of the CSB are also discussed.
\end{abstract}

\newpage
\baselineskip20pt

The Wiedemann-Franz (WF) law is one of the basic properties of Landau's Fermi-liquid theory. It states that at very low temperatures, where elastic collisions dominate, quasiparticles have the same ability to transport heat and charge, leading to a universal relation between the thermal conductivity $\kappa$ and the electrical conductivity $\sigma$, $\frac {\kappa}{\sigma T}=\frac {\pi ^{2}}{3} \left( \frac {k_{B}}{e} \right )^{2}$. Although this law has been shown to be valid in a wide range of materials, recently Hill $et$ $al.$ [1] reported a violation of the WF law in the normal state of a copper-oxide superconductor Pr$_{2-x}$Ce$_{x}$CuO$_{4-y}$ (PCCO) at $T \rightarrow 0$. Later, it was verified that the violation happens only in the underdoped and optimally doped region [2]. In particular, while the charge transport of PCCO at low temperatures is that of a fairly good metal, no residual linear term for the thermal conductivity is observed. This unrelationship between the charge and heat transporting behavior, together with other experiments [3], strongly suggest that the spin and charge degrees of freedom may be separated [1,4] in underdoped cuprate.

At low temperatures, scanning tunneling microscopy (STM) experiments [5] observed a pseudogap in the vortex core which scales as the superconducting (SC) gap. While it is difficult to account for the pseudogap inside the core in conventional weak coupling theories (for a review, see Ref.[6]), these experiments can be understood within several spin-charge separation theories [6-8] that take strong correlations into account. This is another evidence for the existence of spin-charge separation. As the magnetic field increases, the vortices begin to overlap and eventually suppress the superconductivity completely at $H_{c2}$. In this state, the holons are not Bose condensed and have the same ability to transport charge as that of free electrons in metals. Their contribution to the thermal conductivity $\kappa$ at $T \rightarrow 0$ is proportional to $T^{3}$ and hence can be neglected. The spinons all form stable pairs, which prevent single spinons from moving freely. However, the spinon gap of PCCO has a rather pure $d_{x^{2}-y^{2}}$ symmetry [9] and vanishes at the four Fermi points ${\bf k}= \left(\pm \pi/2,\pm \pi/2 \right)$. Owing to this peculiar symmetry, even at $T \rightarrow 0$, a sizeable population of fermions, called nodal fermions, should exist and contribute a linear term to $\kappa$, as in the case of a $d$-wave superconductor [10]. Thus, spin-charge separation alone can not explain the violation of the WF law. A residual linear term for the thermal conductivity $\kappa$ still exists after the spin and charge degrees of freedom are separated.

In this paper we propose that the above inconsistency can be overcomed by the inherent chiral symmetry breaking (CSB) in a spin-charge separated system. Once CSB happens, the originally massless nodal fermions acquire a finite mass via strong gauge interaction. This finite mass is actually a gap below which no free fermions can be excited. In fact, due to the confining potential for the massive fermions, the fermions appear only in the form of stable fermion-antifermion pairs, which are composite Goldstone bosons arising from CSB. The low energy excitations are all bosons, which contribute a $\sim T^{3}$ term to the thermal conductivity. Thus there is no residual linear term at $T \rightarrow 0$, in agreement with experiments.

However, at present spin-charge separation in two dimensions is still an issue of great debate. To test whether it exists in underdoped cuprates, a vison-memory effect was proposed by Senthil and Fisher [11]. In a recent experiment [12], no sign of such an effect was observed, which challenges all the spin-charge separation theories that predict the Senthil-Fisher effect [11,13]. Instead of involving in the intricate details of this experiment, in this paper we choose to begin with the staggered flux phase of the SU(2) formulation of the $t$-$J$ model [14-16] because it does not exhibit the vison-memory effect [7] and naturally leads to a stable $hc/2e$ vortex inside which a finite pseudogap exists [7]. In this phase, the spinon gap have a $d_{x^{2}-y^{2}}$ symmetry and vanishes linearly at the Fermi points, so the low energy excitations from this condensate are massless Dirac fermions [17-19]. To describe the underdoped cuprates more quantitatively, the gauge fluctuations must be included. Of the three components of the SU(2) gauge fields, two become massive via the Higgs mechanism and hence are ignored, leaving a massless gauge field $a_{\mu}$ [16]. Thus we finally arrive at an effective two-dimensional model that consists of massless Dirac fermions, charged bosons and a U(1) gauge field. As we will show below, CSB is an inherent phenomenon in this system.

In the absence of the holons, the effective Lagrangian [18,19] is
\begin{equation}
{\cal L}_{F}=\sum_{\sigma=1}^{N}\overline{\psi}_{\sigma} v_{\sigma, \mu} \left( \partial_{\mu }-ia_{\mu} \right) \gamma_{\mu}\psi_{\sigma}.
\end{equation}
The Fermi field $\psi_{\sigma}$ is a $4\times 1$ spinor. The $4 \times 4$ $\gamma _{\mu}$ matrices obey the algebra, $\lbrace \gamma_{\mu},\gamma_{\nu} \rbrace=2\delta_{\mu \nu}$, and for simplicity, we let $v_{\sigma, \mu}=1$ ($\mu,\nu=0,1,2$). The Lagrangian (1) is the (2+1)-dimensional massless QED (QED$_{3}$), which is invariant under chiral transformations $\psi \rightarrow \exp (i\theta \gamma _{3,5})\psi $, with $\gamma _{3}$ and $\gamma _{5}$ two $4 \times 4$ matrices that anticommute with $\gamma _{\mu}$ ($\mu=0,1,2$). A fermion mass term $m\overline{\psi }\psi$ will break the chiral symmetries. One of the most interesting properties of QED$_{3}$ is that the massless fermions may develop a finite mass via the interaction with strong gauge field, called dynamical CSB [20-22]. CSB is a nonperturbative phenomenon and any calculations based on perturbation theory fail to obtain it. The standard approach to this problem is to solve the Dyson-Schwinger (DS) equation for the fermion self-energy. In general, the inverse fermion propagator is written as $S^{-1}(p)=i\gamma \cdot p A \left( p^{2} \right)+\Sigma \left( p^{2} \right)$, $A(p^{2})$ the wave-function renormalization and $\Sigma(p^{2})$ the fermion self-energy. If the DS equation has only trivial solutions, the fermions remain massless and the chiral symmetries are not broken. If the mass function develops a nontrivial solution, the fermions then acquire a finite mass, which breaks the chiral symmetries (As discussed below, this is not always true). Appelquist $et$ $al.$ [20] found that $\Sigma (p^{2})$ can have a nontrivial solution only for $N<32/\pi^{2}$. In their calculations $A(p^{2})$ was simply ignored, which leads to a gauge-dependent critical fermion number. Later, Nash [21] considered the DS equation for $\Sigma(p^{2})/A(p^{2})$, and obtained a gauge invariant $N_{c}$ which is very near to $32/\pi^{2}$. In the case of cuprates, $N=2<N_{c}$; therefore, the massless fermions obtain a dynamically generated mass.

This critical behavior, however, might be changed by the holons. Generally, it is believed that CSB occurs only in the gauge theories those are asymptotically free, such as QCD and QED$_{3}$. But an additional scalar field might destroy the asymptotic freedom and hence the CSB, as the Higgs fields do in the Standard Model. To examine whether CSB exists, we will reanalyze the DS equation in the presence of a scalar field. Unfortunately, the behavior of the holons are at present poorly understood and it is difficult to derive a microscopic theory for the holons; therefore, proper assumptions and approximations should be made. In practice, previous treatments [19] simply neglect the holons and consider the Lagrangian (1) only. In this paper, we assume that the effective Lagrangian for the holons is that of the (2+1)-dimensional scalar QED
\begin{equation}
{\cal L}_{B}=\left|\left(\partial_{\mu}-ia_{\mu}-ieA_{\mu}\right)b\right|^{2}+m\left|b\right|^{2},
\end{equation}
where $A_{\mu}$ is the external gauge potential and $m$ is the mass of holons. $b=(b_{1},b_{2})$ is a doublet of scalar fields [16]. The minimum of ${\cal L}_{B}$ is located at $\langle b \rangle=0$, so the Bose condensation does not happen. Since spin-charge separation is supposed to exist there is no Yukawa-type coupling term.

Next we will show that CSB still happens in the presence of the holons [23] by considering the DS equation for $\Sigma(p^{2})/A(p^{2})$. To obtain a gauge invariant result, we use a nonlocal gauge propagator
\begin{equation}
D_{\mu \nu}(q)=\frac{1}{q^{2}\Pi(q^{2})}\left( \delta_{\mu \nu}-(1-\xi)\frac{q_{\mu}q_{\nu}}{q^{2}}\right)
\end{equation}
with $\xi$ a gauge parameter and $\Pi(q^{2})$ the vacuum polarization. The one-loop vacuum polarization from the massless fermions is $\Pi_{F}(q^{2})=N/8\left|q\right|$. To the lowest order, we assume that the $only$ contribution of the holons to the DS equation is the one-loop correction to the vacuum polarization $\Pi (q^{2})$, and that the $only$ effect of $A_{\mu}$ is to suppress the Bose condensation. Simple Feynman diagram calculation gives
\begin{equation}
\Pi _{B}(q^{2})=\frac{1}{4\pi} \left\{ -\frac{2m}{q^{2}}+ \frac{q^{2}+4m^{2}}{q^{2}\left| q \right|} \arcsin \left( \frac{q^{2}}{q^{2}+4m^{2}}\right)^{1/2}\right\}.
\end{equation}
For calculational convenience, we set $m=0$; then, $\Pi _{B}(q^{2})=1/8\left| q \right|$. Using the fact that the total vacuum polarization $\Pi$ is the sum of $\Pi _{F}$ and $\Pi _{B}$, we have
\begin{equation}
\Pi (q^{2})=\frac{N+1}{8\left| q \right|}.
\end{equation}
This expression is rather simple, so it can lead us to an analytical result. After performing calculations parallel to that presented in [21], we find that $\Sigma(p^{2})/A(p^{2})\propto p^{t}$ with $t(t+1)=-32/3\pi^{2}(N+1)$, from which we obtain a critical fermion number
\begin{equation}
N ^{\prime}_{c}=\frac{128}{3\pi^{2}}-1.
\end{equation}

This critical number $N^{\prime}_{c}$ is gauge invariant since it is independent of $\xi$. It is larger than the physical number $2$, so the DS equation has nontrivial solutions in the presence of holons. However, while CSB is described by nontrivial solutions of the DS equation, $not$ all nontrivial solutions lead to CSB. It is known that the breaking of a chiral symmetry is always accompanied by a Goldstone boson, which is a pseudoscalar bound state composed of a fermion and an antifermion. If CSB happens, there should be a nontrivial solution for the Bethe-Salpeter (BS) equation of this bound state. In addition, the bound state wave function must satisfy a normalization condition, which can be converted to a sufficient and necessary condition [24] for the nontrivial solutions of the DS equation to signal CSB. It gives a constraint on the form of $\Sigma(p^{2})/A(p^{2})$. In the case of QED$_{3}$, the solutions found in [21] satisfy such a condition [24], so do our solutions since they have the same asymptotic form. Therefore, our $\Sigma(p^{2})/A(p^{2})$ does correspond to CSB solutions. As a result, the nodal fermions become massive. This mass provides a finite gap that should be overcomed before free fermions are excited. At the same time, the BS equation develops a nontrivial solution corresponding to a truely bound state which is a fermion-antifermon pair. The absence of free fermions at low energy can also be seen from the potential for fermions, which behaves like [25]
\begin{equation}
V({\bf x})\sim \frac{\ln\left|{\bf x}\right|}{2\pi(1+\Pi(0))}
\end{equation}
in the infrared region. In the symmetric phase, $\Pi(0)\rightarrow \infty$; so there is not a confining potential between the massless fermions. In the CSB phase, when the finite fermion mass is taken into account, $\Pi(0)$ becomes finite, leading to a confining potential for the massive fermions which binds the massivie fermion into stable pairs. This striking result indicates that while the $d$-wave spinon gap vanishes at four nodes, there are no free fermions at low temperatures, in sharp contrast to the case of a $d$-wave superconductor [10]. This can explain why there is not a residual linear term for the thermal conductivity of PCCO at $T \rightarrow 0$.

In terms of spin-charge separation, other possibilities, such as the spinon localization and the existence of a complex order parameter ($d_{x^{2}-y^{2}}+id_{xy}$ or $d_{x^{2}-y^{2}}+is$) for the spinon pairs, may also explain the absence of the residual linear term for the thermal conductivity. However, we believe our proposal is more natural since it is intrinsic and material-independent, while the spinon localization depends on samples and a complex order parameter is inconsistent with experiments [9].

The CBS is generally interpreted as AF long-range order. Correspondingly, the Goldstone bosons associated to CBS are nothing but the spin waves associated to AF order [18,15,26,27], with spin waves the Goldstone bosons. This strongly suggests that the underlying ground state of the normal state of underdoped cuprates is actually an antiferromagnetism. But it is fundamentally different from the N\'eel state at half-filling. At half-filling, due to strong repulsive force the number of charge carriers at every lattice is exactly one and the sample is an insulator. While in the underdoped region, the holons are mobile and the sample has a metal-like electrical conductivity. This implies that AF order is not necessarily tied to insulating behaviour [28]. However, thermal fluctuations in two-dimensional systems are strong enough to rapidly restore the chiral symmetry, so the AF order is not expected to persist in the normal state in weak magnetic fields.

Next we would like to discuss the robustness of our $N^{\prime}_{c}$. First of all, we consider the influence of a finite mass for gauge field $a_{\mu}$. Although Bose condensation is completely suppressed above $H_{c2}$, $a_{\mu}$ may acquire a finite mass $\eta$ via the instanton effect [29]. Moreover, CSB in QED$_{3}$ is known to be an essentially low-energy phenomenon because only in the infrared region the gauge interaction is strong enough to cause fermion condensation. This effective asymptotic freedom requires fermions be apart from each other. But when the instanton effect is significant the massive gauge field is unable to mediate a long range interaction. Using the propagator, $D_{\mu\nu}(q)= \frac{8}{(N+1)(\left|q\right|+\eta)}\left( \delta_{\mu\nu}-\frac{q_{\mu}q_{\nu}}{q^{2}}\right)$, we obtain the following DS equation
\begin{eqnarray}
\Sigma(p^{2})&=&\frac{4}{(N+1)\pi^{2}p}\int dk\frac{k\Sigma(k^{2})}{k^{2}+\Sigma^{2}(k^{2})} \nonumber \\
&&\times \left(p+k-\left|p-k\right|-\eta\ln\left(\frac{p+k+\eta}{\left|p-k\right|+\eta}\right)\right).
\end{eqnarray}
Here, for simplicity, we set $A(p^{2})=1$ (Note this will lead to a gauge-dependent $N_{c}$; however, since this $N_{c}$ is very near to the gauge invariant one [21], equation (8) is expected to be a good approximation). In the strong coupling limit the instanton effect gives $a_{\mu}$ a very large mass, say $\eta\gg\Lambda$ with $\Lambda$ the ultraviolet cutoff. In this limit, the DS equation becomes
\begin{equation}
\Sigma\arctan\left(\frac{N+1}{8\Sigma}\right)=\frac{N+1}{8}\left(1-\pi^{2}\eta\right).
\end{equation}
Obviously, this equation has no physical solutions; thus, a large mass of $a_{\mu}$ can suppress the CSB. The large $\eta$ also causes spin-charge recombination [19]. Fortunately, as we discussed above, extensive experiments suggest that spinons and holons are actually well separated. Based on this, we expect the system is actually in the weak coupling limit, implying a very small $\eta$ if it is indeed present. Then the last term in the kernel of (8) can be dropped safely, leaving a DS equation which has a critical number very near to our $N^{\prime}_{c}$. So the instanton effect does not significantly change our result.

There are several other corrections that could modify $N^{\prime}_{c}$ than $\eta$. In general, $v_{\sigma,1}\ne v_{\sigma,2}$; actually, the ratio between them is larger than $10$. Recently, this issue was addressed [30] in a physically different but mathematically related model. The result is that a weak velocity anisotropy does not change the critical number, which suggests a stability of our conclusion against the velocity anisotropy. In addition, the holons may have a noticeable mass $m$. For very large $m$, $\Pi_{B}\to0$, then the critical number $N_{c}$ maintains its value in the absence of holons. Next we assume $m\sim \left|p\right|$ in (4), now the vacuum polarization from holons becomes $\Pi_{B}\sim 0.1/8\left|p\right|$, leaving $N_{c}$ essentially unchanged. Since CSB is a low energy phenomenon, the integral over the momentum in the DS equation is within a small interval. Therefore, it is very reasonable to conclude that our result is entirely independent of the holon mass, even if we do not know its exact value. Finally, in this paper we consider only the one-loop contribution to the vacuum polarization from fermions and bosons. Higher-order corrections were shown [21,31] have only minor influence on $N_{c}$. But these calculations said nothing about the higher-order corrections from the holons, which will be discussed in the future.

In conclusion, in a spin-charge separated system, we showed that the massless nodal fermions acquire a finite mass, which breaks the chiral symmetries and provides a finite gap that should be overcomed for free fermions to be excited. This implies that while the $d$-wave spinon gap vanishes at the nodes, at $T \rightarrow 0$ there are no free fermions. Thus no residual linear term for the thermal conductivity can be observed. Our result reveals a very interesting underlying ground state for the pseudogap region of underdoped cuprates where AF order coexists with metal-like electrical conductivity. Till now, no such observations are reported. We expect elaborate experiments, including neutron scattering and STM, would determine whether such a state exists.

$Notes$ $added$: After our work was completed, we noted that Houghton $et$ $al.$ [32] also studied the breakdown of the WF law in terms of spin-charge separation. Their work made a mean-field treatment of the $t$-$J$ model in the large-N limit. In our work, however, strong gauge fluctuation plays an essential role and leades to CSB which was not mentioned in Ref.[32]. We believe CSB is necessary in explaining the violation of the WF law because if it does not occur there should be a residual linear term for the thermal conductivity caused by the low energy fermions excited from the nodes of the $d_{x^{2}-y^{2}}$ spinon gap [10].

G.Z.L. thanks P. A. Lee and M. Reenders for helpful communications. This work is supported by National Science Foundation in China No.10175058.


\begin{thebibliography}{99}

\bibitem{} {R. W. Hill $et$ $al.$, Nature (London) {\bf 414}, 711 (2001).}

\bibitem{} {C. Proust, $et$ $al.$, cond-mat/0202101.}

\bibitem{} {For a review, see P. A. Lee, Physica C {\bf 317-318}, 194 (1999).}

\bibitem{} {P. W. Anderson, Science {\bf 235}, 1196 (1987). }

\bibitem{} {Ch. Renner $et$ $al.$, Phys. Rev. Lett. {\bf 80}, 3606 (1998); S. H. Pan $et$ $al.$, Phys. Rev. Lett. {\bf 85}, 1536 (2000).}

\bibitem{} {M. Franz and Z. Te\v{s}anovi\'{c}, Phys. Rev. B {\bf 63}, 064516 (2001).}

\bibitem{} {P. A. Lee and X.-G. Wen, Phys. Rev. B {\bf 63}, 224517 (2001).}

\bibitem{} {T. Senthil and M. P. A. Fisher, Phys. Rev. B {\bf 63}, 134521 (2001).}

\bibitem{} {C. C. Tsui and J. R. Kirtley, Phys. Rev. Lett. {\bf 85}, 182 (2000).}

\bibitem{} {A. C. Durst and P. A. Lee, Phys. Rev. B {\bf 62}, 1270 (2000), and references therein.}

\bibitem{} {T. Senthil and M. P. A. Fisher, Phys. Rev. Lett. {\bf 86}, 292 (2001).}

\bibitem{} {D. A. Bonn $et$ $al.$, Nature (London) {\bf 414}, 887 (2001).}

\bibitem{} {M. Franz and Z. Te\v{s}anovi\'{c}, Physics C {\bf 357}, 49 (2001).}

\bibitem{} {I. Affleck, Z. Zou, T. Hsu, and P. W. Anderson, Phys. Rev. B {\bf 38}, 745 (1988).}

\bibitem{} {E. Dagotto, E. Fradkin, and A. Moreo, Phys. Rev. B {\bf 38}, 2926 (1988).}

\bibitem{} {X.-G. Wen and P. A. Lee, Phys. Rev. Lett. {\bf 76}, 503 (1996); P. A. Lee, N. Nagaosa, T. K. Ng, and X.-G. Wen, Phys. Rev. B {\bf 57}, 6003 (1998).}

\bibitem{} {J. B. Marston and I. Affleck, Phys. Rev. B {\bf 39}, 11538 (1989); L. B. Ioffe and A. I. Larkin, Phys. Rev. B {\bf 39}, 8988 (1989).}

\bibitem{} {D. H. Kim and P. A. Lee, Ann. Phys. (N.Y.) {\bf 272}, 130 (1999).}

\bibitem{} {W. Rantner and X.-G. Wen, Phys. Rev. Lett. {\bf 86}, 3871 (2001).}

\bibitem{} {T. Appelquist, D. Nash, and L. C. R. Wijewardhana, Phys. Rev. Lett. {\bf 60}, 2575 (1988).}

\bibitem{} {D. Nash, Phys. Rev. Lett. {\bf 62}, 3024 (1989).}

\bibitem{} {E. Dagotto, J. B. Kogut, and A. Koci\'{c}, Phys. Rev. Lett. {\bf 62}, 1083 (1989).}

\bibitem{} {Kim and Lee [17] have discussed the effect of the holons on $N_{c}$. In the presence of the holons they adopted a particular gauge propagator, $D_{\mu \nu}(q)=\frac{1}{q^{2}\Pi _{F}(q^{2})}\delta_{\mu i}\delta_{\nu j} \left( \delta_{ij}-\frac{q_{i}q_{j}}{{\bf q}^{2}}\right)$, and showed that the critical number reduces to $N_{c}/2=16/\pi^{2}<2$ and hence CSB no longer occurs. However, this result is suspectable because whether CSB happens actually depends on the gauge choice once such gauge propagator is used. If we work in another gauge, $D_{\mu \nu}(q)=\frac{1}{q^{2}\Pi_{F}(q^{2})}\delta_{\mu \nu}$, then in the presence of the holons the system acquires a critical number $32/\pi^{2}$, larger than the physical number $2$.}

\bibitem{} {G. Cheng, In $Precision$ $Test$ $of$ $Standard$ $Model$ $and$ $New$ $Physics$, CCAST-WL Workshop Series {\bf 55}, 73 (1996), ed., Chao-hsi Chang; G. Cheng and T. K. Kuo, J. Math. Phys. {\bf 38}, 6119 (1997); G.-Z. Liu and G. Cheng, Phys. Lett. B {\bf 510}, 320 (2001).}

\bibitem{} {P. Maris, Phys. Rev. D {\bf 52}, 6087 (1995).}

\bibitem{} {J. B. Marston, Phys. Rev. Lett. {\bf 64}, 1166 (1990); C. Mudry and E. Fradkin, Phys. Rev. B {\bf 49}, 5200 (1994); R. B. Laughlin, cond-mat/9802180.}

\bibitem{} {I. F. Herbut, Phys. Rev. Lett. {\bf 88}, 047006 (2002); Z. Te\v{s}anovi\'{c}, O. Vafek, and M. Franz, Phys. Rev. B {\bf 65}, 180511 (2002).}

\bibitem{} {L. Balents, M. P. A. Fisher, and C. Nayak, Phys. Rev. B {\bf 60}, 1654 (1999).}

\bibitem{} {A. M. Polyakov, Nucl. Phys. B {\bf 120}, 429 (1977).}

\bibitem{} {D. J. Lee and I. F. Herbut, cond-mat/0201088.}

\bibitem{} {P. Maris, Phys. Rev. D {\bf 54}, 4049 (1996); V. Gusynin, A. Hams, and M. Reenders, Phys. Rev. D {\bf 63}, 045025 (2001).}

\bibitem{} {A. Houghton, S. Lee, and J. B. Marston, Phys. Rev. B {\bf 65}, 220503 (2002).}
\end{thebibliography}
\end{document}